# Breaking the Fixed-Substrate Limitation of Deep Learning Ellipsometry with Pseudo-Substrate Spectra

Yuki Ikarashi[1], Naoki Kurokawa[1], James N. Hilfiker[2], Takuji Maekawa[3], Hitoshi Tampo[4], and Hiroyuki Fujiwara[1,a)]

[1]*Department of Electrical, Electronic and Computer Engineering, Gifu University, 1-1 Yanagido, Gifu, 501-1193, Japan*
[2]*J.A. Woollam Co., Inc., 311 South 7th Street, Lincoln, Nebraska, 68508, USA*
[3]*Rohm Research and Development Center, ROHM Co., Ltd., 21, Saiin Mizosaki-cho, Ukyo-ku, Kyoto, 615-8585, Japan*
[4]*Research Institute for Energy Efficient Technologies, National Institute of Advanced Industrial Science and Technology (AIST), Central2, 1-1-1 Umezono, Tsukuba, Ibaraki, 305-8568, Japan*

a) Author to whom correspondence should be addressed: fujiwara.hiroyuki.u7@f.gifu-u.ac.jp

## ABSTRACT

A machine learning method, deep learning ellipsometry (DLE), enables direct and rapid estimation of the optical constants and structures of thin films from spectroscopic ellipsometry (SE) spectra. However, conventional neural-network-based DLE models are restricted to the specific substrate used during training, and applying them to a different substrate requires ~100 h of retraining. Here, we propose a framework that overcomes this fixed-substrate limitation by introducing a two-input feature-wise linear modulation DLE (FiLM-DLE). In this model, the SE spectra of a thin-film structure are input together with those of the substrate measured before film deposition, which we term the pseudo-substrate spectra. Any substrate, including a multilayer stack, is thus represented solely by its measured SE response, analogous to the pseudo-dielectric function, so that no prior knowledge of the substrate structure is required. Trained on a virtual dataset of 165 million samples representing metal, glass, and $SiO_2$/Si substrates with different $SiO_2$ thicknesses, a single FiLM-DLE model, trained once, extracts thin-film optical constants and structures with high accuracy and shows strong applicability to substrates not included in the training datasets. Experimental validation on 47 thin-film samples across diverse substrates yielded sample-averaged mean absolute errors, relative to conventional manual analysis, of 0.047 for refractive index, 0.030 for extinction coefficient, and 1.52 nm for thickness, with each analysis completed in under 2 s. FiLM-DLE thus eliminates the need for a separate DLE model for each substrate and enables rapid, high-accuracy optical characterization of thin films on diverse substrates.

The introduction of artificial intelligence (AI) based on neural networks has significantly advanced automated spectroscopic ellipsometry (SE) analysis, enabling rapid, high-throughput characterization with reduced analyst dependence.[1-10] More recently, various deep learning ellipsometry (DLE) approaches have been reported for the direct estimation of thin-film thicknesses, optical constants [i.e., refractive index ($n$) and extinction coefficient ($k$)], and dielectric function parameters from SE spectra.[1-5,9] Other methods utilize deep learning to classify material types from SE heatmaps.[8] Most recently, we proposed a highly accurate, general-purpose DLE method based on a convolutional neural network (CNN) that automatically determines the optical constants, thicknesses, and bandgap $E_g$ of various semiconductor films in less than 2 s.[6]

However, DLE models reported to date have generally been restricted to the specific substrate assumed in the training dataset or optical model.[1-9] In our previous studies,[6,8] for example, the DLE model was trained assuming a fixed $SiO_2$(2 nm)/Si substrate. Consequently, this model essentially cannot be applied to thin films formed on other substrates, such as glass and metals, because the interference effects differ significantly depending on the substrate. Moreover, our general-purpose DLE method requires 100 million simulated structures and ~100 h for data generation and model training, making retraining for each substrate impractical. Thus, the fixed-substrate limitation remains a major obstacle to the practical application of DLE. Although an AI-based approach for real-time in situ monitoring of absorbing films on unknown substrates has been reported,[10] a general DLE framework for ex situ analysis of thin films on arbitrary substrates has not been established.

To overcome the fixed-substrate limitation and broaden the applicability of DLE, we have developed an algorithm that incorporates the SE spectra of the substrate structure, hereafter referred to as the pseudo-substrate spectra, as a second input in addition to the thin-film SE spectra. In particular, we leveraged feature-wise linear modulation (FiLM), a general conditioning technique developed for visual question answering, in which intermediate image features are modulated by a language input.[11] By replacing the language input with the pseudo-substrate spectra, we constructed a two-input FiLM-DLE capable of analyzing single-layer thin films on diverse substrates without requiring separate models.

**Fig. 1** shows the overall architecture of the proposed two-input FiLM-DLE model. In this model, Input A is the SE spectra of a single-layer thin-film structure based on a surface roughness/bulk layer/substrate optical model. Input B is the pseudo-substrate spectra ($\psi_{sub}$, $\Delta_{sub}$), i.e., the SE spectra measured for the substrate before thin-film deposition: regardless of its actual composition and layer structure, the underlying structure is represented solely by its measured SE response, in the same manner as the pseudo-dielectric function.[12] This treatment is analogous to the virtual substrate approximation in real-time SE analysis,[13,14] in which an arbitrary underlying structure is replaced by an effective substrate derived from measured ($\psi$, $\Delta$). Accordingly, Si substrates with different $SiO_2$ overlayer thicknesses, for example, are incorporated directly through their ($\psi_{sub}$, $\Delta_{sub}$) spectra without any prior knowledge of the layer structure. Unlike the virtual substrate approximation, however, FiLM-DLE never converts ($\psi_{sub}$, $\Delta_{sub}$) into an effective dielectric function or uses them in an optical calculation; they serve only as a conditioning input that identifies the substrate, whereas the training spectra are calculated exactly from the full layer structure. Accordingly, no restriction is imposed on the film thickness (10–300 nm in this study), and the applicable range is determined by the substrate structures covered in the training data.

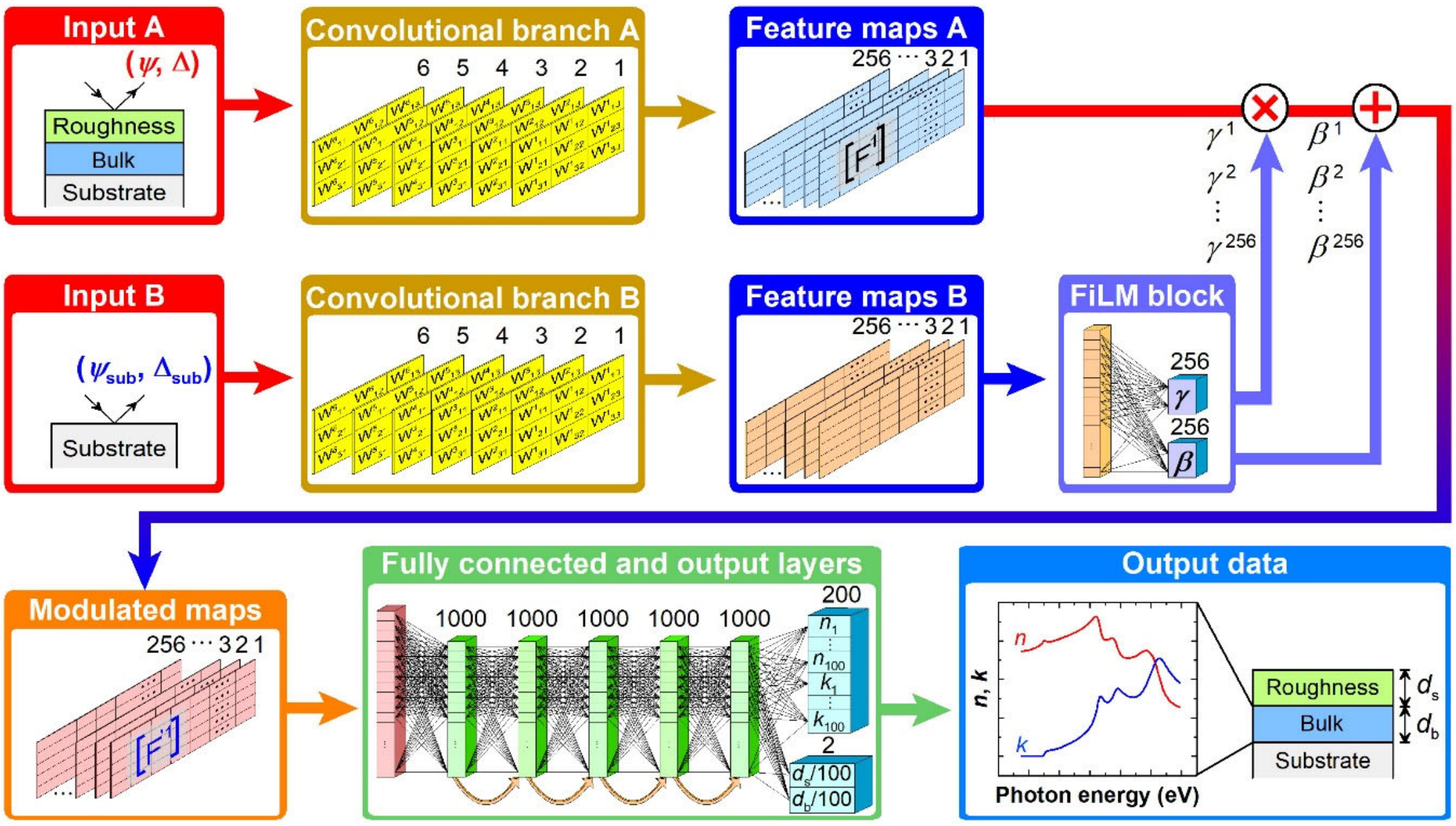

**FIG. 1.** Overall architecture of the proposed two-input feature-wise linear modulation deep learning ellipsometry (FiLM-DLE) model. Input A is the SE spectra of the thin-film structure ($\psi$, $\Delta$), whereas Input B is the pseudo-substrate spectra ($\psi_{sub}$, $\Delta_{sub}$), i.e., the SE spectra of the substrate structure measured before film deposition.

Each input is then processed by a convolutional branch of six layers with 3×3 kernels, yielding 256 feature maps (see **Fig. S1** in the Supplementary Material); Feature maps A represent the SE response of the entire thin-film structure, whereas Feature maps B represent the SE response originating from the substrate structure.

Feature maps B are then passed through the FiLM block, which generates a scaling coefficient $\gamma^c$ and a shift $\beta^c$ for each channel ($c$ = 1, 2, …, 256). These coefficients modulate the corresponding feature map in Feature maps A; specifically, each element $F_{(i,j)}^{c}$ is transformed as follows:[11]

$$F_{(i,j)}^{\prime c} = \gamma^c F_{(i,j)}^{c} + \beta^c \qquad (1)$$

Here, $F_{(i,j)}^{\prime c}$ denotes the corresponding element of the modulated map. Through this operation, the features extracted from the thin-film SE spectra are conditioned on the substrate-specific optical response embedded in the pseudo-substrate spectra, allowing FiLM-DLE to account for substrate-induced variations in the SE response.

The resulting tensor is flattened and passed through five fully connected (FC) layers, each containing 1000 nodes, with skip connections. The output layer predicts 100-point $n$ and $k$ spectra, along with the surface roughness layer thickness $d_s$ and the bulk layer thickness $d_b$. ReLU activation and batch normalization are applied to all hidden layers (see **Fig. S1** in the Supplementary Material).

Training data were generated by forward calculations based on the Fresnel equations, following our previous DLE study.[6] For all datasets, the SE spectra were calculated at five angles of incidence (50°, 55°, 60°, 65°, and 70°) and 100 photon energies over the range of 0.75–6.0 eV. Accordingly, each SE dataset is 1000-dimensional [2 ($\psi$, $\Delta$) × 5 (AoI) × 100 (photon energy)], whereas the target vector is 202-dimensional [2 ($n$, $k$) × 100 (photon energy) + 2 ($d_s$, $d_b$)]. The $d_b$ and $d_s$ were varied uniformly over ranges of 10–300 nm and 0–10 nm, respectively. The optical constants of the thin-film bulk layer were described by 1–10 Tauc–Lorentz (TL) oscillators.[15] Each TL parameter was varied randomly within a realistic range based on an established materials database.[16]

Using these common thin-film conditions, we generated three large-scale datasets (Datasets 1–3) for thin films on metal, glass, and $SiO_2$/Si substrates. All models in this study share the FiLM-DLE architecture of **Fig. 1** and differ only in their training data: three individual models, Metal-FiLM (Dataset 1), Glass-FiLM (Dataset 2), and $SiO_2$/Si-FiLM (Dataset 3), each applicable to one substrate group, and Unified-FiLM, trained on Datasets 1–3 combined and applicable to all.

For Dataset 1 (metal substrates), the optical constants of the metal substrates ($n_{sub}$, $k_{sub}$) were described by a Drude term[17] and 2–6 TL terms.[15] This procedure yielded 100 million training samples (see **Fig. S2** in the Supplementary Material for examples of the generated ($n_{sub}$, $k_{sub}$) spectra). For Dataset 2 (glass substrates), the substrate optical constants were modeled using a Cauchy dispersion model[18] and a Gaussian model.[19] A total of 40 million training samples were generated (see Eqs. (S1) and (S2) in the Supplementary Material). For Dataset 3 ($SiO_2$/Si), the optical constants of Si and $SiO_2$ were fixed,[16] while the $SiO_2$ layer thickness $d_o$ was varied randomly over the range of 0–50 nm, and a total of 25 million training samples were generated. In the case of Unified-FiLM, a total of 165 million samples obtained by combining Datasets 1–3 were used.

Generation of all three training datasets (165 million samples in total) using an Intel Core i9-14900KF CPU required 64 h 21 min. To optimize the model, the loss function was defined as the simple sum of the mean-squared errors (MSEs) for ($n$, $k$) and ($d_s$/100, $d_b$/100). Training was performed with the Adam optimizer on an NVIDIA GeForce RTX 4090 GPU (24 GB) and required 72 h 14 min for Unified-FiLM.

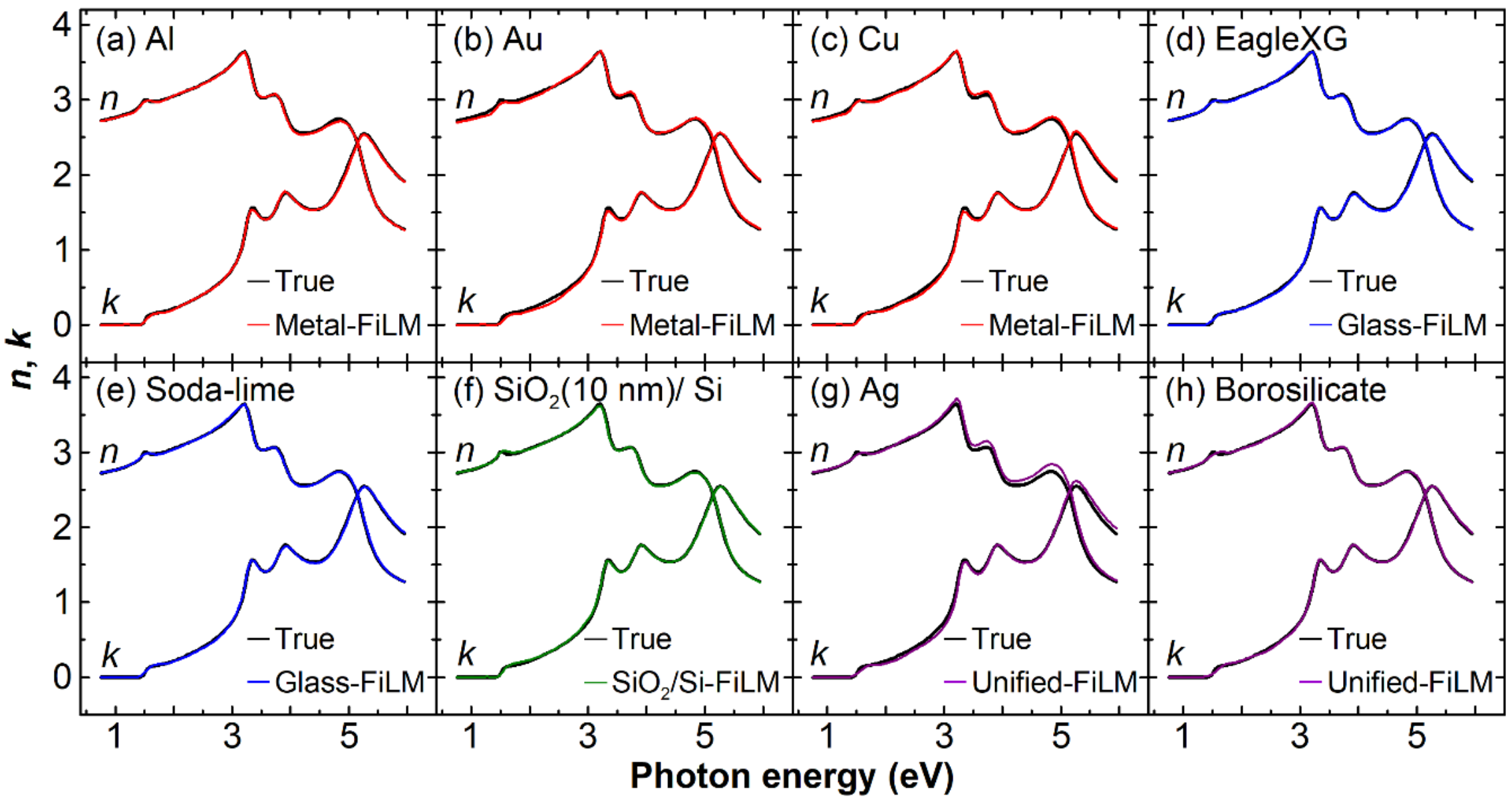

**FIG. 2.** Predicted $n$ and $k$ spectra for a simulated CdTe thin film ($d_s$ = 7 nm, $d_b$ = 135 nm) formed on (a) Al, (b) Au, (c) Cu, (d) Eagle XG glass, (e) Soda-lime glass, (f) $SiO_2$(10 nm)/Si, (g) Ag, and (h) borosilicate glass substrates. The black lines show the true spectra, whereas the red, blue, green, and purple lines show the predictions from Metal-FiLM, Glass-FiLM, $SiO_2$/Si-FiLM, and Unified-FiLM, respectively.

We first examined whether the fixed-substrate limitation could be overcome without FiLM. The prediction accuracy of the conventional DLE model decreases markedly even when the $SiO_2$ layer thickness differs from the training condition by as little as 5 nm (see **Fig. S3** in the Supplementary Material). We also trained the conventional single-input DLE model, which has no substrate input, on the combined Datasets 1–3, i.e., the same data as used for Unified-FiLM. This model yields substantially larger prediction errors (see **Fig. S4** in the Supplementary Material), indicating that the substrate information must be supplied explicitly.

**Figs. 2(a)-2(f)** show the *n* and *k* spectra predicted by the individual FiLM-DLE models (Metal-FiLM, Glass-FiLM, and $SiO_2$/Si-FiLM) for a simulated CdTe thin film on Al, Au, Cu, Eagle XG, Soda-lime glass, and $SiO_2$(10 nm)/Si substrates. Despite the substantial differences in substrate optical properties, the predicted spectra agree almost perfectly with the true spectra under all substrate conditions, confirming the high prediction accuracy of each model for its substrate group.

Notably, Unified-FiLM, trained on the combined Datasets 1–3, maintains prediction accuracy equivalent to that of the individual models [**Figs. 2(g)** and **2(h)** for Ag and borosilicate glass substrates, respectively]. Thus, a single FiLM-DLE model can replace substrate-specific models.

To further evaluate the substrate generalization of Unified-FiLM, we performed a simulation study under 37 substrate conditions (**Fig. 3**). For each substrate, 107–123 simulated thin-film materials were evaluated using six different thickness combinations (see the **GitHub repository** linked in the Supplementary Material). Prediction accuracy was evaluated using the mean absolute error (MAE), averaged over all predictions for each substrate:

$$y_{\mathrm{ave}}^{\mathrm{MAE}} = \frac{1}{m}\sum_{r=1}^{m}\left[\frac{1}{p}\sum_{q=1}^{p}\left|y_{r,q} - \hat{y}_{r,q}\right|\right] \qquad (2)$$

In Eq. (2), $p$, $m$, $y_{r,q}$, and $\hat{y}_{r,q}$ are the number of photon-energy points, the number of predictions, the true value, and the predicted value, respectively. For the Al substrate, for example, 123 materials and six thicknesses yielded $m$ = 738 predictions, with $n_{\mathrm{ave}}^{\mathrm{MAE}}$=0.017 and $k_{\mathrm{ave}}^{\mathrm{MAE}}$=0.0095, respectively.

**Fig. 3** summarizes the prediction accuracy of simulated thin films ($n_{\mathrm{ave}}^{\mathrm{MAE}}$ and $k_{\mathrm{ave}}^{\mathrm{MAE}}$) obtained for different substrate materials using Unified-FiLM (closed circles). For all transparent, $SiO_2$/Si, and metal substrates, we observed consistently low values of $n_{\mathrm{ave}}^{\mathrm{MAE}}$ and $k_{\mathrm{ave}}^{\mathrm{MAE}}$ below 0.04. In particular, the transparent substrates yielded extremely small $n_{\mathrm{ave}}^{\mathrm{MAE}}$ values of ~0.001, reflecting the small variation in their optical constants. Importantly, the consistently low errors for $SiO_2$/Si substrates with various $SiO_2$ thicknesses demonstrate that FiLM-DLE accurately accounts for substrate structures containing an overlayer. In contrast, the errors for the metal substrates increase slightly because their optical constants vary strongly with photon energy. For the trained substrates (i.e., transparent, $SiO_2$/Si, and metal substrates), the average errors were $n_{\text{ave}}^{\text{MAE}}$=0.015 and $k_{\text{ave}}^{\text{MAE}}$=0.014.

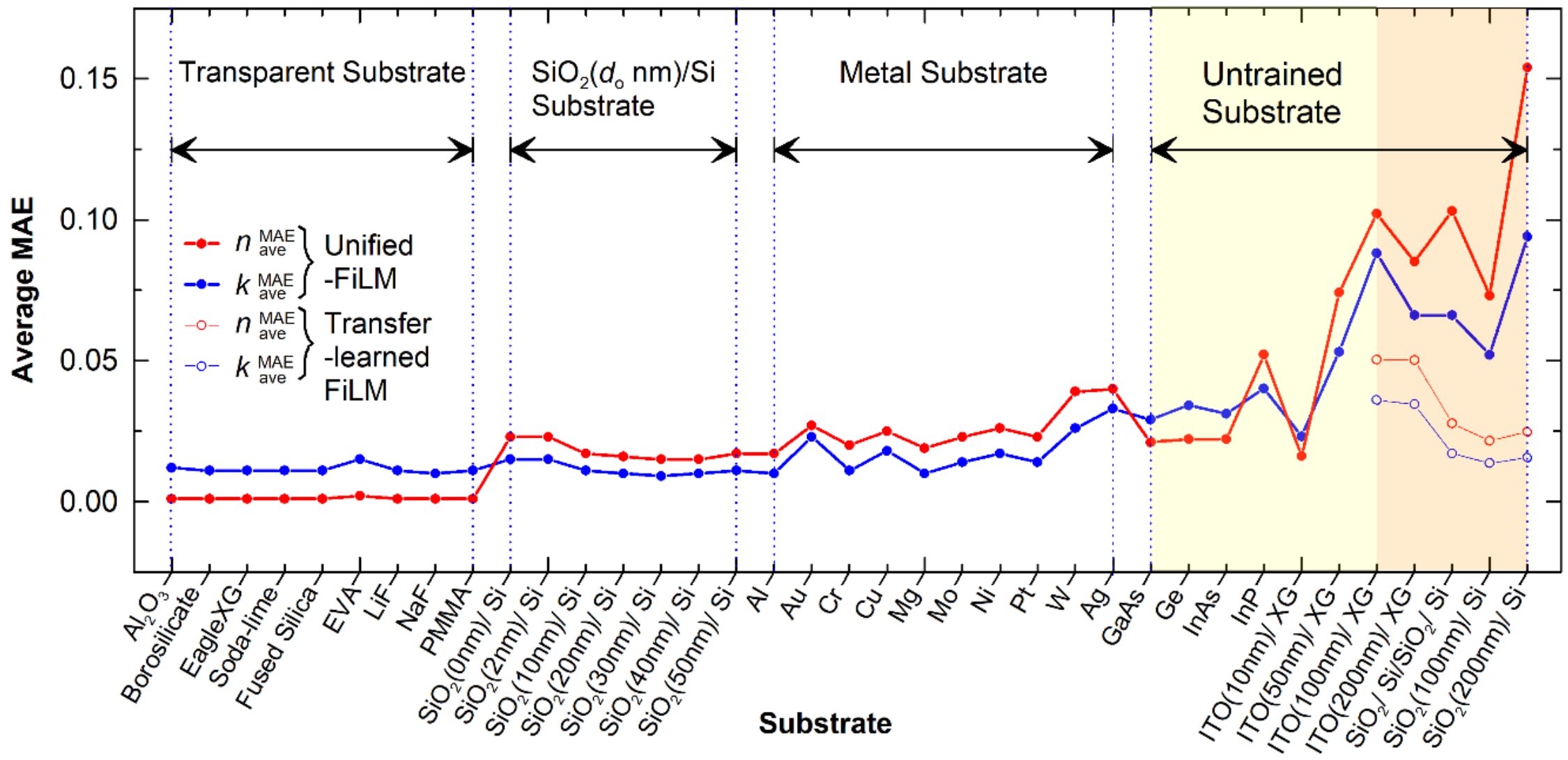

**FIG. 3.** Substrate-generalization performance of Unified-FiLM for 37 substrate conditions. Closed circles show the results of the model trained on transparent substrates, $SiO_2$/Si substrates, metal substrates, and substrates not included in the training (highlighted in yellow and orange). Open circles show the results after transfer learning, in which the substrates in the orange area were added to the training. Each point is the MAE averaged over more than 100 thin-film materials with six different thicknesses. XG denotes Eagle XG glass.

Remarkably, Unified-FiLM also showed high prediction accuracy for the untrained substrates highlighted by the yellow background in **Fig. 3**. In particular, when the ($\psi_{\text{sub}}$, $\Delta_{\text{sub}}$) spectra calculated for the ITO/Eagle XG glass stack were used as the pseudo-substrate spectra, small prediction errors were obtained even for a multilayer structure of surface roughness/bulk layer/ITO($\leq$ 50 nm)/Eagle XG. In contrast, as highlighted by the orange background in **Fig. 3**, when the thicknesses of overlayers on substrates are large ($\geq$ 100 nm) or a substrate has a complex multilayer structure, the prediction accuracy degrades rather notably. This degradation reflects the limited range of substrate structures in the training datasets rather than an intrinsic limitation of the method. Indeed, when we retrained the Unified-FiLM model by transfer learning including these substrate structures, low analysis errors were obtained, as indicated by the open circles in **Fig. 3**, demonstrating that Unified-FiLM is readily extended to substrates with complex structures.

Furthermore, the thickness prediction errors were evaluated separately for $d_b$ and $d_s$ using the following average absolute error:

$$\Delta d_{\text{ave}} = \frac{1}{m}\sum_{r=1}^{m}\left|d_r - \hat{d}_r\right| \qquad (3)$$

In Eq. (3), $m$ is the total number of predictions, and $d_r$ and $\hat{d}_r$ are the true and predicted thicknesses for prediction $r$, respectively. Applying Eq. (3) to the simulated thin films on the trained substrates separately for $d_b$ and $d_s$ yielded $\Delta d_{\text{b,ave}}$=1.35 nm and $\Delta d_{\text{s,ave}}$=0.55 nm, respectively. These results confirm the high generalization performance of Unified-FiLM across diverse substrate optical constants, thin-film materials, and thickness conditions using a single model.

**Fig. 4** shows the prediction performance of Unified-FiLM for experimental thin-film samples including an organic thin film (Spiro-TTB) formed on Al, a microcrystalline Si ($\mu$c-Si) film formed on a float glass substrate, and a $CuInGaSe_2$ (CIGS) film formed on a $SiO_2$(2 nm)/Si substrate. Note that the measured SE spectra were denoised using a general-purpose machine-learning denoising model,[9] because spectral noise seriously degrades the inference accuracy.[6,9] In **Fig. 4**, the $n$ and $k$ spectra predicted by Unified-FiLM show excellent agreement with the results of conventional manual SE analysis based on the B-spline model.[20] The maximum absolute thickness errors relative to the manual analysis were $\Delta d_b$=1.9 nm and $\Delta d_s$=0.3 nm. Furthermore, for the $\mu$c-Si/float glass sample, replacing the pseudo-substrate spectra with those of an Al substrate markedly reduced the prediction accuracy, confirming that Unified-FiLM indeed exploits the pseudo-substrate information (see **Fig. S5** in the Supplementary Material).

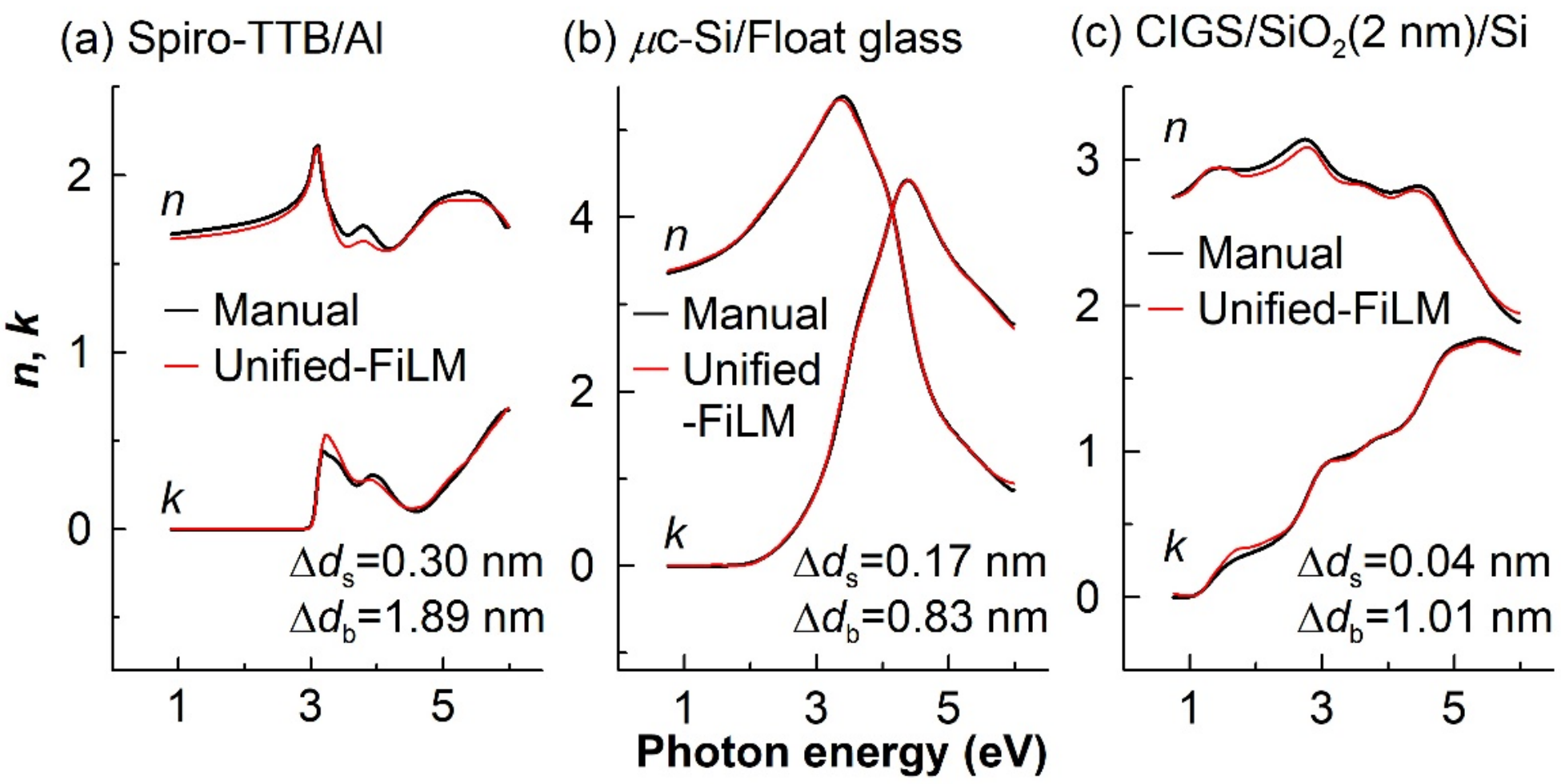

**FIG. 4.** Predicted optical constants $n$ and $k$ for experimental thin-film samples obtained using Unified-FiLM: (a) Spiro-TTB/Al, (b) $\mu$c-Si/float glass, and (c) CIGS/$SiO_2$(2 nm)/Si. The black lines show the results of manual SE analysis using the B-spline model,[20] whereas the red lines show the predictions obtained using Unified-FiLM. The absolute thickness errors, $\Delta d_s$ and $\Delta d_b$, are also given in each panel.

To evaluate performance on experimental samples more comprehensively, we analyzed 47 thin films: 2 on metal, 18 on glass, and 27 on $SiO_2$(2 nm)/Si substrates, covering oxide, semiconductor, and organic films (see **Table SI** in the Supplementary Material). **Table I** summarizes the results obtained using the individual FiLM-DLE models and Unified-FiLM, together with those of our conventional substrate-specific DLE model (base model) for the $SiO_2$/Si samples. Here, $k_{\mathrm{ave}}^{\mathrm{MAE}}$, $n_{\mathrm{ave}}^{\mathrm{MAE}}$, $\Delta d_{\mathrm{b,ave}}$, and $\Delta d_{\mathrm{s,ave}}$ were calculated from Eqs. (2) and (3) by taking the results of the manual SE analysis as the true values. For the metal-substrate samples, Unified-FiLM yielded larger$n_{\mathrm{ave}}^{\mathrm{MAE}}$, $k_{\mathrm{ave}}^{\mathrm{MAE}}$, and $\Delta d_{\mathrm{s,ave}}$ than Metal-FiLM with similar $\Delta d_{\mathrm{b,ave}}$, although this comparison is based on only two samples. For the glass-substrate samples, in contrast, Unified-FiLM reduced the errors notably relative to Glass-FiLM, presumably owing to the larger and more diverse combined training dataset, whereas for the $SiO_2$/Si samples it maintained accuracy comparable to that of $SiO_2$/Si-FiLM. Thus, Unified-FiLM retains practical accuracy while extending its applicable range to all three substrate groups.

For all 47 samples, the average prediction errors of Unified-FiLM were $k_{\mathrm{ave}}^{\mathrm{MAE}}$=0.030, $n_{\mathrm{ave}}^{\mathrm{MAE}}$=0.047, $\Delta d_{\mathrm{b,ave}}$=1.52 nm, and $\Delta d_{\mathrm{s,ave}}$=0.67 nm, which are comparable to those of the conventional base model specialized for $SiO_2$(2 nm)/Si substrates. For the $SiO_2$(2 nm)/Si samples themselves, the errors of Unified-FiLM are slightly larger than those of the base model, which was trained with 100 million samples for this single substrate condition, whereas the $SiO_2$/Si group accounts for only 25 million of the 165 million samples used for Unified-FiLM and covers $SiO_2$ thicknesses of 0–50 nm. Nevertheless, Unified-FiLM achieved practical accuracy using a single model while substantially expanding the applicable substrate range, and its inference time remained below 2 s per sample, as in conventional DLE. Moreover, the 72-h training of Unified-FiLM (**Table I**) is a one-time cost, whereas conventional DLE requires ~100 h for data generation and retraining for each new substrate.

**TABLE I.** Analysis results obtained using the individual FiLM-DLE models, Unified-FiLM, and conventional base model for SE data measured from 47 experimental samples.

| **Model** | **Evaluation Material** | $k_{\mathrm{ave}}^{\mathrm{MAE}}$ | $n_{\mathrm{ave}}^{\mathrm{MAE}}$ | $\Delta d_{\mathrm{b,ave}}$ (nm) | $\Delta d_{\mathrm{s,ave}}$ (nm) | **Training Data (million)** [a)] | **Training Time** |
|---|---|---|---|---|---|---|---|
| Metal-FiLM | Metal | 0.022 | 0.030 | 2.24 | 0.38 | 100 | 44 h 26 min |
| Glass-FiLM | Glass | 0.020 | 0.037 | 2.15 | 0.96 | 40 | 9 h 07 min |
| $SiO_2$/Si-FiLM | $SiO_2$(2nm) | 0.037 | 0.062 | 1.64 | 0.97 | 25 | 11 h |
| Unified-FiLM | Metal | 0.050 | 0.063 | 1.78 | 0.61 | 165 | 72 h 14 min |
| | Glass | 0.013 | 0.021 | 1.05 | 0.29 | | |
| | $SiO_2$(2nm) | 0.040 | 0.064 | 1.81 | 0.93 | | |
| | Average | 0.030 | 0.047 | 1.52 | 0.67 | | |
| base model[b)] | $SiO_2$(2nm) | 0.023 | 0.040 | 1.04 | 0.62 | 100 | 53 h 50 min |

(a) Number of training samples used for model training.

(b) Conventional DLE model trained using a fixed Si substrate covered with a native oxide layer.[6]

In summary, we have developed a two-input FiLM-DLE framework that fundamentally overcomes the fixed-substrate limitation of conventional deep learning ellipsometry by explicitly incorporating the pseudo-substrate spectra, i.e., the SE spectra of the substrate structure measured prior to film deposition, into the deep-learning process. A single Unified-FiLM model achieved rapid (<2 s) and accurate analysis of 47 experimental thin-film samples on optically diverse substrates, with optical-constant errors comparable to those of conventional substrate-specific DLE.

## Acknowledgement

This work was supported by JSPS KAKENHI Grant Number 23K03923. We gratefully acknowledge Dr. Takuya Matsui of National Institute of Advanced Industrial Science and Technology (AIST) for generously providing experimental samples essential to this study.

## Conflict of Interest

The authors have no conflicts to disclose.

## Author Contributions

Yuki Ikarashi: Conceptualization(equal), Formal analysis(equal), Writing- review & editing(equal). Naoki Kurokawa: Formal analysis(equal). James N. Hilfiker: Provide experimental ellipsometry data and samples, Formal analysis(equal). Takuji Maekawa: Provide semiconductor samples. Hitoshi Tampo: Provide compound semiconductor samples. Hiroyuki Fujiwara: Conceptualization(equal), Supervision, Writing- review & editing(equal).

## Data availability

The simulation data used for the evaluation in Fig. 3 are available in the GitHub repository linked in the supplementary material. Other data are available from the corresponding author upon reasonable request.

**SUPPLEMENTARY MATERIAL**

**Breaking the Fixed-Substrate Limitation of Deep Learning Ellipsometry with Pseudo-Substrate Spectra**

Yuki Ikarashi[1], Naoki Kurokawa[1], James N. Hilfiker[2], Takuji Maekawa[3], Hitoshi Tampo[4], and Hiroyuki Fujiwara[1]

[1]*Department of Electrical, Electronic and Computer Engineering, Gifu University, 1-1 Yanagido, Gifu, 501-1193, Japan*
[2]*J.A. Woollam Co., Inc., 311 South 7th Street, Lincoln, Nebraska, 68508, USA*
[3]*Rohm Research and Development Center, ROHM Co., Ltd., 21, Saiin Mizosaki-cho, Ukyo-ku, Kyoto, 615-8585, Japan*
[4]*Research Institute for Energy Efficient Technologies, National Institute of Advanced Industrial Science and Technology (AIST), Central2, 1-1-1 Umezono, Tsukuba, Ibaraki, 305-8568, Japan*

## I. Generation of training data for glass substrates

For Dataset 2 (glass substrates), the optical constants of the glass substrates ($n_{sub}$, $k_{sub}$) were represented using a Cauchy–Schott-type dispersion model[1] and a Gaussian model,[2] as given by

$$(n_{sub})^2 = A' + \frac{B'}{\lambda^2} + \frac{C'}{\lambda^4} - D'\lambda^2 \qquad \text{(S1)}$$

$$k_{sub} = K_0 \exp\left[-\frac{(E - E_0)^2}{2\sigma^2}\right] \qquad \text{(S2)}$$

Here, $\lambda$ is the wavelength in nm, and $E$ is the photon energy in eV. $A'$, $B'$, $C'$, and $D'$ are the Cauchy–Schott parameters, whereas $K_0$, $E_0$, and $\sigma$ represent the intensity, peak position, and peak width of the Gaussian model, respectively.

## II. Thin-film materials used in the simulation evaluation

The simulation data for the thin-film materials used in the comprehensive evaluation shown in **Fig. 3** are available in the following **GitHub repository**:

https://github.com/fujiwaralab-720/Fujiwara-net

The repository contains the optical constants of 123 thin-film materials listed in the materials database[3] in CSV format. All 123 materials were used for the evaluations on the metal and $SiO_2$/Si substrates, whereas 107 materials were used for the evaluation on the transparent substrates. For each thin-film material, six randomly selected combinations of the surface roughness layer thickness $d_s$ and bulk layer thickness $d_b$ were used.

**TABLE SI.** Thin-film materials and substrates of the 47 experimental samples analyzed in this study. Each entry is written as thin-film material(number of samples)/substrate. The number in parentheses immediately after each thin-film material indicates the number of samples with that film/substrate combination.

| Substrate group | Material (number of samples)/ Substrate |
|---|---|
| Metal | Spiro-TTB(1)/ Al, a-Si:H(1)/ Al |
| Glass | Organic(1)/ Fused silica |
| | Spiro-TTB(1)/ Eagle XG, a-Si:H(1)/ Eagle XG |
| | $\mu$c-Si(2)/ Float glass, a-Si(1)/ Float glass |
| | $SiN_x$(2)/ Glass, Si(1)/ Glass, a-Si(1)/ Glass, $WO_3$(1)/ Glass, ZnSnO(1)/ Glass, NiCrO(1)/ Glass, $Si_3N_4$(2)/ Glass, a-Si:H(1)/ Glass, SiN(1)/ Glass, Polyimide(1)/ Glass |
| $SiO_2$/Si | a-Si:H(1)/ $SiO_2$(2 nm)/Si, a-SiO:H(6)/ $SiO_2$(2 nm)/Si, $CuInGaSe_2$(7)/ $SiO_2$(2 nm)/Si, $Cu_2ZnGeSe_4$(4)/ $SiO_2$(2 nm)/Si, $Cu_2Zn(Sn,Ge)Se_4$(5)/ $SiO_2$(2 nm)/Si, $PbI_2$(3)/ $SiO_2$(2 nm)/Si, ZnO(1)/ $SiO_2$(2 nm)/Si |

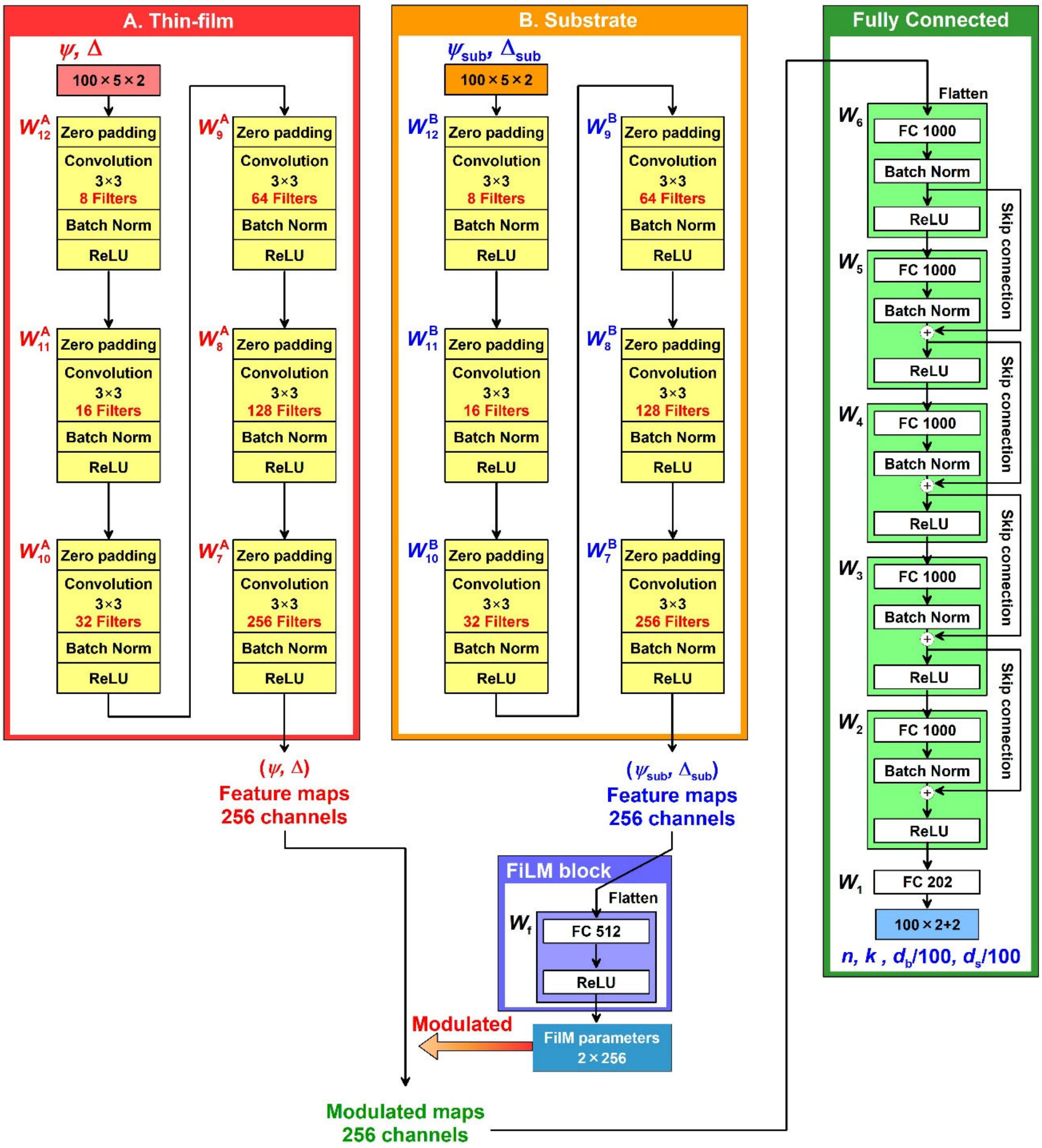

**FIG. S1.** Detailed block architecture of the two-input FiLM-DLE model shown in **Fig. 1**. The red and orange blocks represent the convolutional branches that process Input A, the SE spectra ($\psi$, $\Delta$) of the thin-film structure, and Input B, the SE spectra ($\psi_{sub}$, $\Delta_{sub}$) of the corresponding substrate, respectively. The purple FiLM block generates FiLM parameters from Feature maps B and applies them to modulate Feature maps A. The resulting tensor is flattened and passed through the green Fully

Connected block. The output layer predicts the 100-point $n$ and $k$ spectra, along with the surface roughness layer thickness $d_s$ and the bulk layer thickness $d_b$.

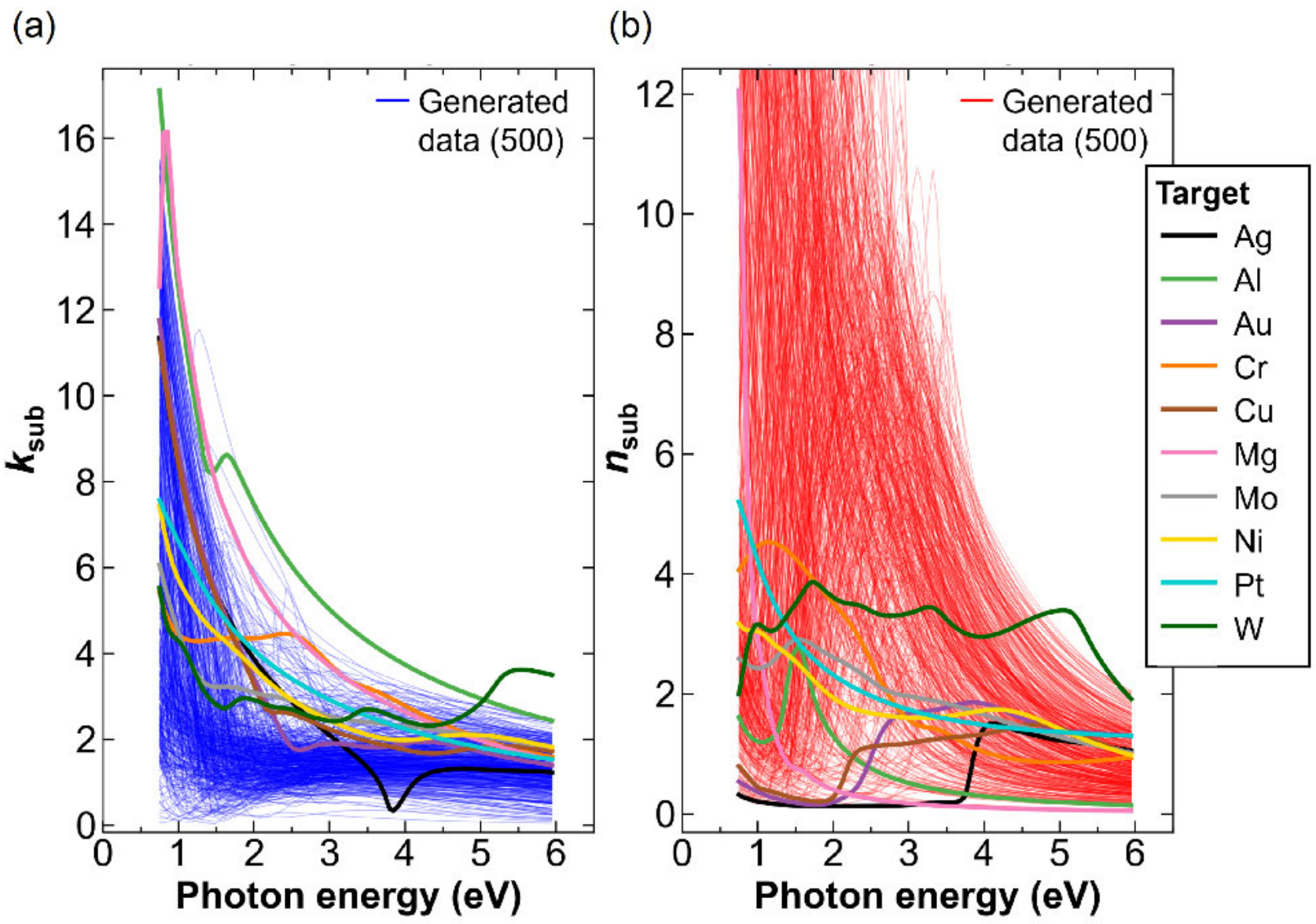

**FIG. S2.** Representative optical constants randomly generated for the metal substrates in Dataset 1: (a) Extinction coefficient $k_{sub}$ and (b) refractive index $n_{sub}$. The blue lines in (a) and red lines in (b) show 500 optical constants randomly generated by varying the parameters of the Drude[4] and Tauc–Lorentz[5] terms. The colored solid lines show the optical constants of 10 reference metal substrates (Target) obtained from the materials database.[3]

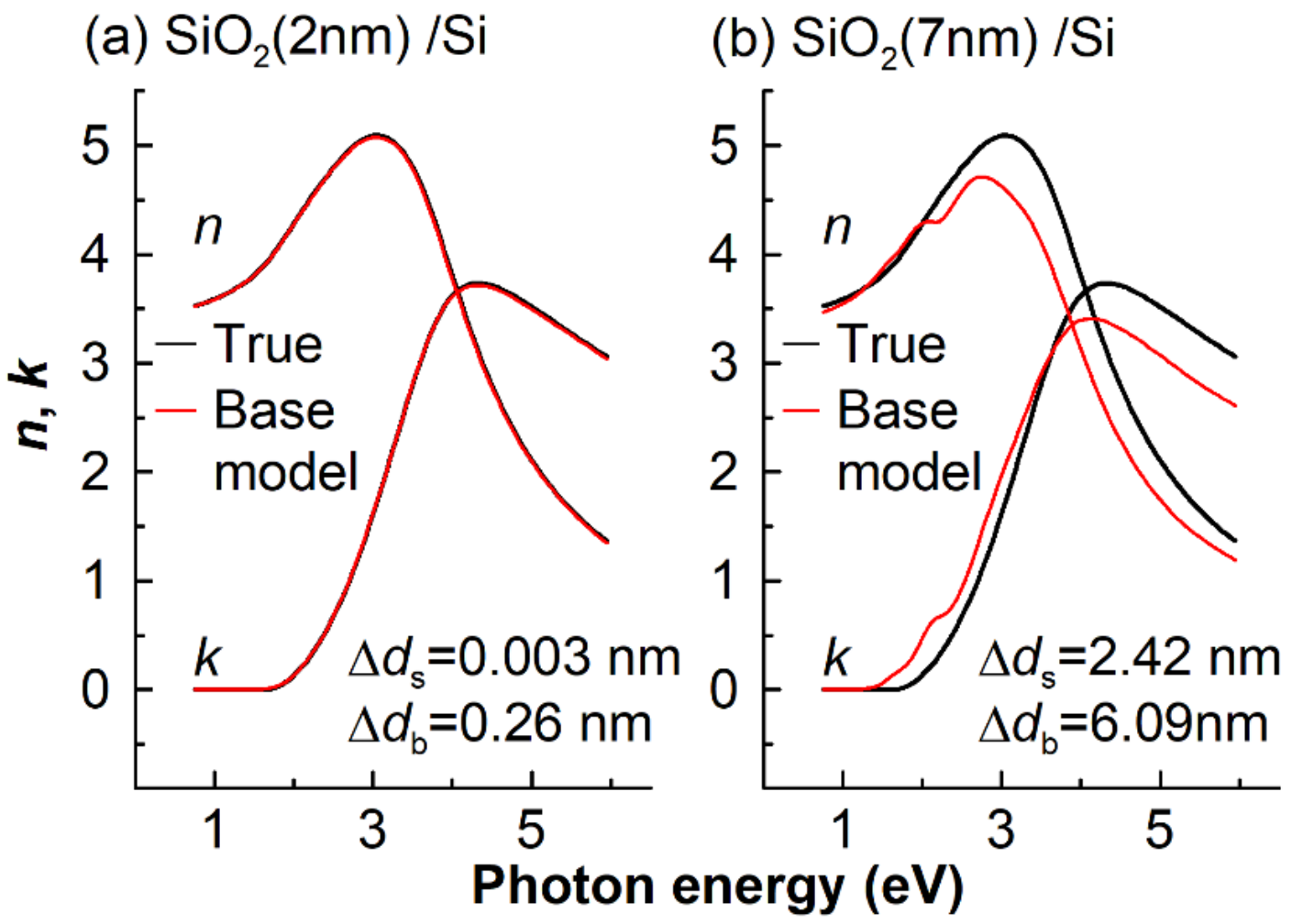

**FIG. S3.** Prediction results obtained using the conventional base model[6] for a simulated a-Si:H thin film formed on (a) $SiO_2$(2 nm)/Si and (b) $SiO_2$(7 nm)/Si substrates. The base model was trained using data generated assuming a fixed $SiO_2$(2 nm)/Si substrate, and the same a-Si:H thin film was simulated in both cases. In (b), the $SiO_2$ layer in the simulated structure is 7 nm thick, i.e., 5 nm thicker than in the training condition. The black lines show the true *n* and *k* spectra used in the simulation, whereas the red lines show the predictions obtained using the base model. The absolute thickness errors $\Delta d_s$ and $\Delta d_b$ are also shown in each panel.

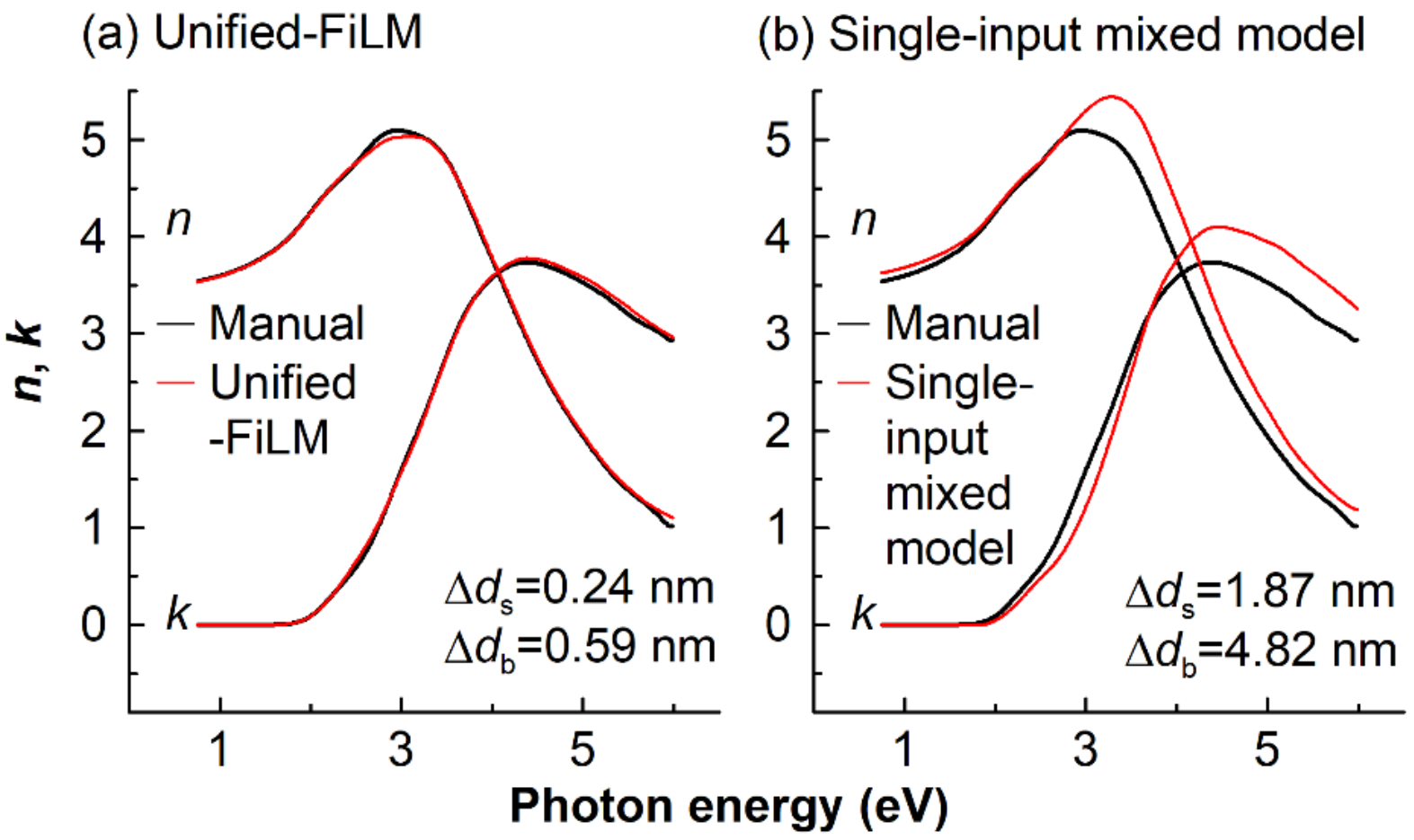

**FIG. S4.** Analysis results for the measured SE spectra of an a-Si thin film formed on a float glass substrate. The predictions obtained using (a) Unified-FiLM and (b) a single-input mixed model, i.e., the conventional single-input architecture of the base model[6] (no substrate input) trained on the same combined Datasets 1–3 as Unified-FiLM, are shown. The black lines show the optical constants *n* and *k* obtained by manual SE analysis using the B-spline model,[7] whereas the red lines show the predictions obtained using each model. The absolute thickness errors $\Delta d_s$ and $\Delta d_b$ are also shown in each panel.

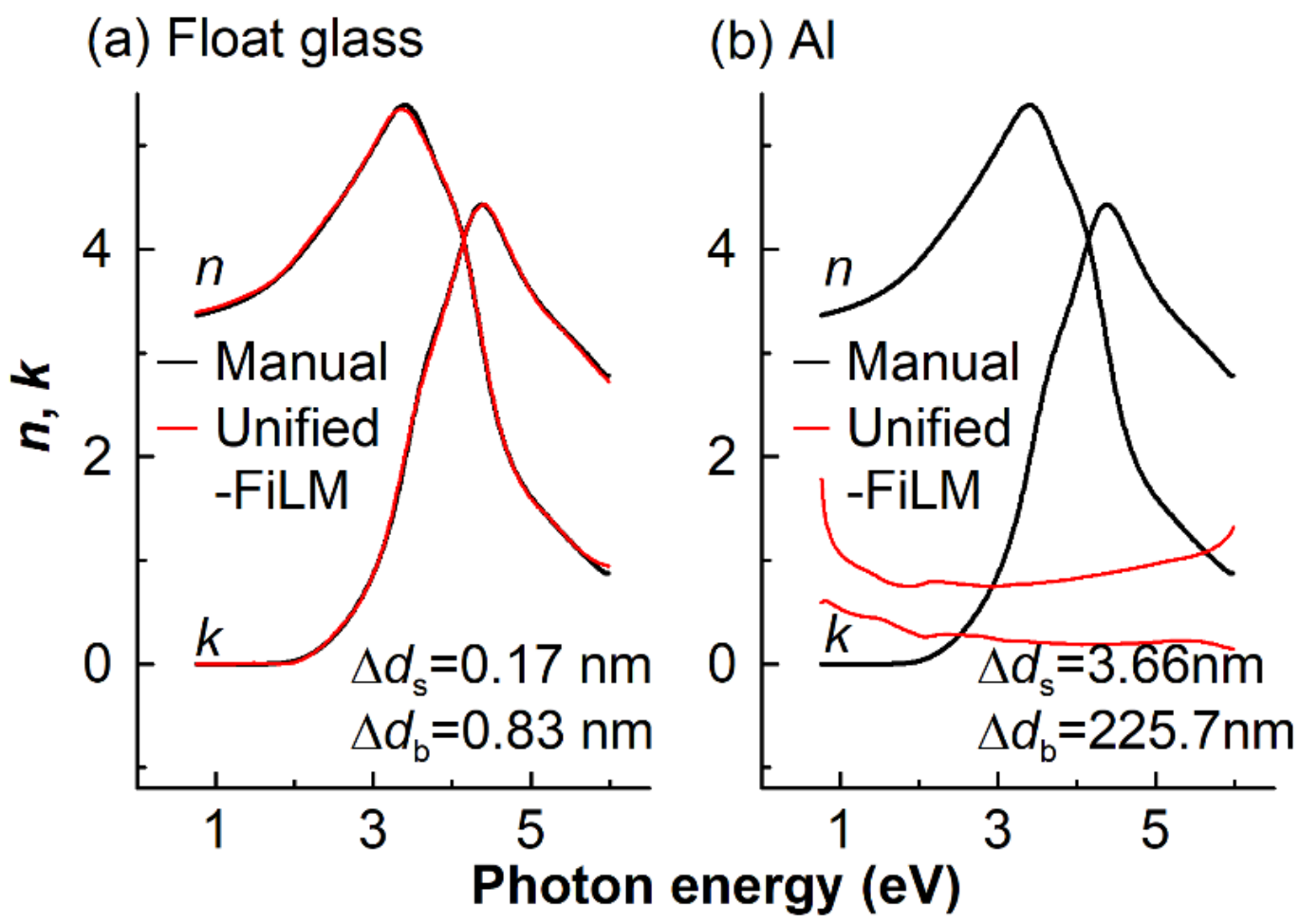

**FIG. S5.** Analysis results obtained using Unified-FiLM for an experimental $\mu$c-Si thin film formed on a float glass substrate. Panels (a) and (b) show the results obtained using the SE spectra of the actual float glass substrate and an Al substrate different from the actual deposition substrate, respectively, as the substrate input ($\psi_{sub}$, $\Delta_{sub}$). In both cases, the same measured SE spectra ($\psi$, $\Delta$) of the $\mu$c-Si/float glass thin film were used as Input A. The black lines show the results of manual SE analysis using the B-spline model,[7] whereas the red lines show the predictions obtained using Unified-FiLM. The absolute thickness errors $\Delta d_s$ and $\Delta d_b$ are also shown in each panel.